\documentclass[aps,twocolumn]{revtex4}
\usepackage{longtable,dcolumn}
\usepackage{epsfig,epsf,color,colordvi}
\usepackage{amsfonts,amssymb,amsmath,amsthm,bm}
\usepackage{pifont} 

\renewcommand{\thesection}{\arabic{section}}
\renewcommand{\thesubsection}{\Alph{subsection}}
\newcommand{\mysection}[1]{\refstepcounter{section}\section*{\large\thesection\hspace*{2ex}\normalsize #1}}
\newcommand{\mysubsection}[1]{\refstepcounter{subsection}\subsection*{\thesection.\thesubsection\hspace*{3ex} #1}}

\numberwithin{equation}{section}

\theoremstyle{remark} 
\newtheorem{thm}{Theorem}[section] 
\newtheorem{lem}[thm]{Lemma}
\newtheorem{prop}[thm]{Proposition}

\theoremstyle{remark}

\newtheorem{exmp}{Example}[section] 
\newtheorem{numberedrem}{Remark}[section]

\theoremstyle{remark} 
\newtheorem*{rem}{Remark}
 
\newtheorem*{notn}{Notation} 
\newtheorem*{minorexmp}{Example}

\newcommand{\lsup}[2]{{\vphantom{#2}}^{#1}{#2}}

\newcommand{\etal}{~\mbox{{\it et al.\/}}}

\newcommand{\dash}{\mbox{\,---\,}}

\newcommand{\newtable}{\begin{table}}                
\newcommand{\newfigurewide}{\begin{figure*}}         
\newcommand{\newfigure}{\begin{figure}}              

\newcommand{\bA}{A}
\newcommand{\ztwo}{\mathbb{Z}_2}
\newcommand{\soplus}[1]{\underset{#1}{\oplus}}
\newcommand{\sotimes}[1]{\underset{#1}{\otimes}}
\newcommand{\lstar}{\text{\ding{83}}}  
\newcommand{\fR}{\mathbb{R}}
\newcommand{\fC}{\mathbb{C}}

\newcommand{\fH}{\mathbb{H}}
\newcommand{\fF}{\mathbb{F}}
\newcommand{\fA}{\mathbb{A}}

\begin{document}

\title{
$\mathbf{\ztwo^n}$ grading of the classical Lie algebras.

}
\author{P. E. Maslen}\email{maslen@rutgers.edu}
\affiliation{
Rutgers University\\
Camden, NJ 08102, USA.
}
\begin{abstract}
The $\ztwo^n$ gradings of the classical Lie algebras are described.
To elucidate the grading,
the classical Lie algebras are expressed in terms of matrix algebras, $\text{mat}(N;\fF)$, 
where $\fF$ is one of eight fields or Clifford algebras which carry gradings
ranging from zero to $\ztwo^3$.  When the matrix dimension $N$ is divisible by a
power of two, the Lie algebras have higher gradings which may be expressed in terms of higher
Clifford algebras: these are also described.
\end{abstract} 
\keywords{} 
\maketitle

\mysection{Introduction}
In Weyl's original classification of the classical Lie algebras~\cite{weyl46},
scant regard was paid to their graded structure.
The introduction of supersymmetry and Lie superalgebras to
physics~\cite{berezin2,haag1,kac1} led to renewed scrutiny of the
properties of Lie algebras concomitant with their grading, and in $1978$
Rittenberg, Wyler and Lukierski discovered the generalized Lie algebras and 
superalgebras, also
known as Lie color algebras and Lie color superalgebras~\cite{rittenberg1,rittenberg2,lukierski1}.  
Rittenberg \etal\ 
demonstrated how to construct a Lie color algebra 
from a Lie algebra with a grading, and Scheunert
subsequently proved that all Lie color algebras can
be obtained in this manner~\cite{scheunert78}.

A knowledge of the grading of a Lie algebra is required for the construction of the associated
Lie color algebras: this provided the
impetus for the present study of the $\ztwo^n$ gradings of the classical Lie algebras.

The paper is organized as follows.  Sec.~\ref{s.intro} contains introductory remarks on graded
algebras and Clifford algebras and also serves to define the notation. 

Sec.~\ref{s.la_f} describes a graded classification of the classical Lie algebras, based on 
their description in terms of matrix algebras over one of eight graded fields or Clifford algebras.

Sec.~\ref{s.la_r} describes higher gradings of Lie algebras, for example the grading of
$O(N;\fR)$ when $N$ is divisible by a power of two, and recasts these Lie algebras in terms of matrix algebras over
higher Clifford algebras.  Sec.~\ref{s.conclusion} presents concluding remarks.

\mysection{Conventions on graded algebras}\label{s.intro}

Throughout this work $\Gamma$
denotes a finite {\em abelian\/} group.  
The groups of principal relevance are $\ztwo$ and $\ztwo^n$, the direct sum
of $n$ copies of $\ztwo$.  A convenient representation of $\ztwo$ is the
set $\{0,1\}$ with addition modulo $2$, and the corresponding representation 
of $\ztwo^n$ consists of the $2^n$ $n$-tuples comprised of $1$'s and $0$'s.

A {\em $\mathit{\Gamma}$-graded vector space} consists of a vector space $V$ equipped
with a partition into subspaces, each of which is labelled by an
element of $\Gamma$,
\[
V = \soplus{\gamma\in\Gamma} V_\gamma
\]

A {\em $\mathit{\Gamma}$-graded algebra} consists of an algebra $\bA$ 
which is a graded vector space,
\begin{gather*}
\bA = \soplus{\gamma\in\Gamma} \bA_\gamma \\
\intertext{and additionally satisfies,}
\bA_\alpha \bA_\beta  \subset  \bA_{\alpha+\beta},\qquad\forall\, \alpha,\beta \in\Gamma  
\end{gather*}
It follows that if $\bA$ has a unit element then it belongs to $\bA_0$.

Consider a graded algebra equipped with an adjoint, denoted by $\dag$.  In addition to the usual
conditions for an anti-involution,
\begin{align}
(x y)^\dagger & =  y^\dagger x^\dagger\qquad\forall x,y\in A\nonumber\\
(x^\dagger)^\dagger & =  x \nonumber\\
\intertext{the adjoints employed in this work are also required to preserve the grading,}
(A_\alpha)^\dag & \subset  A_\alpha \label{eq.preserve}
\end{align}

The {\em Clifford algebra $Cl(d_+,d_-;\fR)$} is a real algebra of dimension $2^{\,d_+ +d_-}$,
generated by $d_+ +d_- $ basis elements $e_i$ satisfying,
\begin{align*}
e_i^2 &  = 
\begin{cases}
 1, & 1 \le i \le d_+ \\ 
-1, & d_+ < i \le d_+ +d_-
\end{cases}\\
e_i e_j + e_j e_i & = 0,\qquad  i \ne j \\
e_1 e_2\ldots e_{d_+ +d_-}  & \ne  \pm 1
\end{align*}
$e_{ijk}$ is a shorthand notation for $e_i e_j e_k$.
Each Clifford algebra has several adjoint operators, collectively denoted by $\dagger$. 
The first adjoint,
{\em Clifford Reversion}, reverses the order of the Clifford product,
\begin{subequations}
\begin{gather}
\widetilde{e_i e_j e_k} := e_k e_j e_i = -e_{ijk}
\label{eq.rev} \\
\intertext{while {\em Clifford Conjugation} also introduces a sign-change,}
\overline{e_i e_j e_k} := (-e_k) (-e_j) (-e_i) = e_{ijk}
\label{eq.cliff}
\end{gather}
Additional adjoints can be obtained by changing the sign of selected $e_i$'s:
\begin{notn} $\displaystyle x^{\bm{\dag(a_+,a_-)}}$ denotes the adjoint obtained by composing Clifford Reversion with the following sign-change, 
\begin{align}
\left(e_i e_j e_k\right)^{\dag(a_+,a_-)} & :=  e_k^{\dag(a_+,a_-)} e_j^{\dag(a_+,a_-)} e_i^{\dag(a_+,a_-)} \nonumber \\
e_i^{\dag(a_+,a_-)} & :=  
\begin{cases}
 e_i,  &   1       \le i \le a_+ \\ 
 -e_i, & a_+       < i \le d_+ \\ 
 e_i,  & d_+       < i \le d_+ + a_- \\ 
 -e_i, & d_+ + a_- < i \le d_+ + d_- 
\end{cases}
\label{eq.adm}
\end{align}
\end{notn}
\label{eq.anti}
\end{subequations}
Thus $x^{\dag(0,0)} = \bar{x}$ and $x^{\dag(d_+,d_-)} = \tilde{x}$.

$Cl(d_+,d_-;\fR)$ has a natural $\ztwo^{d_+ +d_-}$ grading.  The grade associated with
$e_{ijk}$ is the $(d_+ +d_-)$-tuple with $1$'s in positions $i,j,k$
and $0$'s in the remaining positions.
$e_{ijk}^\dag$ has the same grade as $e_{ijk}$, as required by Eq.~\eqref{eq.preserve}.
\begin{minorexmp}$e_{124} \in Cl(5,0;\fR)$ has grade $(1,1,0,1,0)$. \end{minorexmp}

If algebras $A$ and $B$  have gradings $\ztwo^{n_a}$ and $\ztwo^{n_b}$,  
then the tensor product $A\otimes B$ has a $\ztwo^{n_a+n_b}$ grading,
\begin{equation}
\bA_\alpha \otimes B_\beta =: (\bA\otimes B)_{\alpha \oplus \beta}
\label{eq.tp_grade}
\end{equation}

For the purposes of grading, all algebras are regarded as real algebras.  Hence Clifford
algebras over the complex field, $Cl(d;\fC)$, are regarded as tensor products
of real algebras,
\begin{subequations}
\begin{gather}
Cl(d_+ + d_-;\fC) \cong \fC \otimes_\fR Cl(d_+,d_-;\fR)
\intertext{where $\fC \cong Cl(0,1;\fR)$ has a $\ztwo^1$ grading.
Adjoints act on $Cl(1;\fC)$ as,}
\left(\fC \otimes_\fR Cl(d_+,d_-;\fR)\right)^\dag = \fC \otimes_\fR Cl(d_+,d_-;\fR)^\dag
\end{gather}
Adjoints which include complex conjugation can also be defined, but are not employed in this work.
\label{eq.complex}
\end{subequations}

Consider the {\em matrix algebra $\fF(D)$} of $D\times D$ matrices over $\fF$,
where $\fF$ is a Clifford algebra.  $\fF(D)$ may be regarded as a real tensor
product,
\begin{subequations}
\begin{equation}
\fF(D) \cong \fR(D) \otimes_\fR \fF
\label{eq.tp} 
\end{equation}
and it follows that $\fF(D)$ inherits the grading of $\fF$.

The adjoint of the tensor-product $\fF(D)$ is the tensor product of the
adjoints for $\fR(D)$ and $\fF$,
\begin{gather}
\left(
\fR(D) \otimes_\fR \fF
\right)^\dagger
= \fR(D)^\dagger \otimes_\fR \fF^\dagger
\label{eq.tp_ad} \\
\intertext{
and conforms with Eq.~\eqref{eq.preserve}.
The adjoints for $\fF$ are given by Eq.~\eqref{eq.anti}.
The adjoint of $\fR(D)$ is~\cite{porteous95},
}
\fR(D)^\dagger = g^{-1} \fR(D)^t g
\label{eq.r_ad}
\end{gather}
\label{eq.tp_ad_def}
\end{subequations}
where $\fR(D)^t$ denotes the matrix transpose and $g\in \fR(D)$ is the metric.
$\fR(D)$ admits adjoints with both skew-symmetric and symmetric non-degenerate metrics.
The symmetric metrics may be either Euclidean, $g = I$,
or pseudo-Euclidean with signature $(D_+,D_-)$, $g = I_{D_+,D_-}$. 
\begin{flushleft}
\begin{raggedright} \begin{tabular}{lp{0.32\textwidth}}
{\em Notation.} & \\
 ${\fF(D_+,D_-)}$ &   \mbox{$\fF(D_+ +D_-)$ equipped with $g = I_{D_+,D_-}$} \\
 ${(\fF(D_+,D_-),\dag)}$   & $\fF(D_+,D_-)$ equipped with an adjoint, cf Eq.~\eqref{eq.tp_ad} \\
 ${\widetilde{\fF}(D_+,D_-)}$   & $(\fF(D_+,D_-),\dag)$ where $\fF^\dag = \tilde{\fF}$. \\
 ${\overline{\fF}(D_+,D_-)}$   & $(\fF(D_+,D_-),\dag)$ where $\fF^\dag = \bar{\fF}$. \\
\end{tabular}
\end{raggedright}
\end{flushleft}

\begin{table*} 
\begin{minipage}{0.75\textwidth}
\caption{
\mbox{Classification of the Classical Lie Algebras.\protect\footnote{
A blank in the table indicates that the corresponding Lie algebra is superfluous to
the classification.
}}
}\label{t.la}
\begin{tabular}{rlrrlrl} 
\hline\hline
\multicolumn{2}{c}{$\fF$}  & $\ztwo^n$        & 
\multicolumn{4}{c}{$\fF^\dagger$~(Eq.'s~\protect\eqref{eq.tp_ad_def}~and~\protect\eqref{eq.la_def}) 

}  \\
 \cline{4-7} 
&    & \text{grading} & \multicolumn{2}{c}{Clifford Reversion} & \multicolumn{2}{c}{Clifford Conjugation} \\ 
\hline
$\fR$ & $\cong Cl(0,0;\fR)$    & $\dash$     & $O(D_+,D_- ;\fR)$ &            & $\dash$ &                    \\
$\fC$ & $\cong Cl(0,1;\fR)$    & $\ztwo^1$ & $O(D ;\fC)$ &             & $U(D_+,D_- ;\fC)$ &               \\
$\lsup{2}{\fR}$ & $\cong Cl(1,0;\fR)$ & $\ztwo^1$ & $\dash$ &                  & $U(D ; \lsup{2}{\fR})$ & $\cong Gl(D;\fR)$ \\
$\fH$ & $\cong Cl(0,2;\fR)$ & $\ztwo^2$ & $O(D ;\fH)$ &$\cong O\lstar(2D;\fC)$ & $U(D_+,D_- ;\fH)$ &$\cong USp(2D_+,2D_- ;\fC)$  \\
$\fH''$ & $\cong Cl(2,0;\fR)$\footnote{$\fH''\cong Cl(2,0;\fR)\cong\fR(2)$}
& $\ztwo^2$ & $\dash$ &                   & $U(D ;\fH'')$ &$\cong Sp(2D;\fR)$ \\
$\lsup{2}{\fC}$ & $\cong Cl(1;\fC)$\footnote{
$\lsup{2}{\fC}\cong Cl(1;\fC) \cong Cl(0,1;\fR)\otimes_\fR Cl(1,0;\fR)$ .
$\;\;U(D ;\lsup{2}{\fC})$ may be obtained using any of the following adjoints:
(i) Clifford Conjugation on $Cl(1;\fC)$ .  This is equivalent to
    Clifford Conjugation on $Cl(1,0;\fR)$ and the Identity operator on $Cl(0,1;\fR)$.
(ii) Clifford Conjugation on both $Cl(1,0;\fR)$ and $Cl(0,1;\fR)$.
}
  & $\ztwo^2$ & $\dash$ & & $U(D ; \lsup{2}{\fC})$ & $\cong Gl(D;\fC)$ \\
$\lsup{2}{\fH}$ & $\cong Cl(0,3;\fR)$ & $\ztwo^3$ & $U(D ;\lsup{2}{\fH})$ & $\cong U\lstar(2D;\fC)$\footnote{
$U(D ;\lsup{2}{\fH})\cong Gl(D;\fH)\cong U\lstar(2D;\fC)$
} & $\dash$ & \\
 $\fC\otimes_\fR\fH''$ & $\cong Cl(3,0;\fR)$\footnote{
$\fC\otimes_\fR\fH'' \cong Cl(3,0;\fR) \cong \fC(2)$
}
& $\ztwo^3$ & $\dash$ & & $O_\fC U_{\fH''}(D; \fC\otimes_\fR\fH'')$
& $\cong Sp(2D;\fC)$  \footnote{
The notation $O_\fC U_{\fH''}$ indicates that Clifford Conjugation on 
$\fC\otimes_\fR\fH''$ is equivalent to the Identity operator on $\fC$ and 
Clifford Conjugation on $\fH''$.  The Identity operator coincides with
Clifford Reversion on $\fC\cong Cl(0,1;\fR)$.
} \\
\hline
\hline
\end{tabular}
\end{minipage}
\end{table*}

\mysection{$\fF$-grading of the classical Lie algebras}\label{s.la_f}
\mysubsection{Graded classification}\label{s.la_f_class}
Each classical Lie algebra (LA) may be represented as a subalgebra of
$\fF(D)$, where $\fF$ is a field or more generally a Clifford algebra.
The LA then inherits the $\ztwo^n$ grading of $\fF$.  The purpose of
this section is thus to find an $\fF$ with the
highest possible grading for each LA.  The classical LA's associated with
each $\fF$ are described
in this section, and the defining equations for each LA are presented
in Sec.~\ref{s.la_def}.

Table~\ref{t.la} lists the $\fF$'s employed in this
work, together with isomorphic Clifford algebras and their $\ztwo^n$ gradings.

\begin{subequations}
The fields,
\begin{equation}
\fF \in \{\fR,\fC\}
\label{eq.f0}
\end{equation}
permit the classical LA's to be classified as~\cite{gilmore74}:
orthogonal $O(D_+,D_- ;\fF)$, unitary $U(D_+,D_- ;\fF)$, symplectic $Sp(2D;\fF)$, general-linear $Gl(D;\fF)$,
unitary-symplectic $USp(2D_+ ,2D_- ;\fC)$, $O\lstar(2D;\fC)$ and $U\lstar(2D;\fC)$.

If Eq.~\eqref{eq.f0} is augmented with the quaternion field,
\begin{equation}
\fF \in \{\fR,\fC\} \cup \{\fH\}
\label{eq.f1}
\end{equation}
then $USp$, $O\lstar$ and $U\lstar$ may be reclassified as unitary, orthogonal and
general-linear algebras over $\fH$~\cite{gilmore74}, and this yields a higher
grading than Eq.~\eqref{eq.f0}.
\begin{minorexmp}
$U\lstar(2D;\fC)\cong Gl(D;\fH)$ and hence admits a $\ztwo^2$ grading,
versus a $\ztwo^1$ grading using Eq.~\eqref{eq.f0}.
\end{minorexmp}

If Eq.~\eqref{eq.f1} is augmented with three Clifford
algebras,
\begin{equation}
\fF \in  \{\fR,\fC,\fH\} \cup 
\{\lsup{2}{\fR}, \lsup{2}{\fC}, \lsup{2}{\fH}\}
\label{eq.f2}
\end{equation}
then the general-linear algebras may be reclassified as unitary~\cite{porteous95},
and this yields a higher grading than Eq.~\eqref{eq.f1}.
\begin{minorexmp}
$Gl(D;H)\cong U(D ;\lsup{2}{\fH})$ and hence admits a $\ztwo^3$ grading, versus a $\ztwo^2$
grading using Eq.~\eqref{eq.f1}.
\end{minorexmp}

The six $\fF$'s in $Eq.~\eqref{eq.f2}$ arise naturally in the context of Clifford
algebras, because every Clifford algebra is isomorphic to $\fF(2^n)$
for one such $\fF$~\cite{cartan1908}.  However, in order to expose the $\ztwo^n$ grading of all
the classical LA's, we have found it necessary to 
supplement $Eq.~\eqref{eq.f2}$ with two
additional Clifford algebras, the split-quaternions $\fH''$ and the 
biquaternions $\fC\otimes_\fR \fH''$,
\begin{equation}
\fF \in  \{\fR,\fC,\fH, \lsup{2}{\fR}, \lsup{2}{\fC}, \lsup{2}{\fH}\}
\cup 
\{\fH'' , \fC\otimes_\fR \fH'' \}
\label{eq.f3}
\end{equation}
\end{subequations}
Eq.~\eqref{eq.f3} permits the symplectic algebras to be reclassified,
thereby yielding a higher grading than Eq.~\eqref{eq.f2}.
\begin{minorexmp}
$Sp(2D;\fR)\cong U(D ;\fH'')$ and hence admits a $\ztwo^2$ grading, versus no
grading using Eq.~\eqref{eq.f2}.
\end{minorexmp}

Table~\ref{t.la} contains the classification of the classical LA's 
with respect to Eq.~\eqref{eq.f3}.  
A blank in the table indicates that
the corresponding LA is superfluous to the classification. 
For instance, neither $O(D ;\fH'')\cong O(2D;\fR) \subset O(D_+,D_- ;\fR)$ nor 
$O(D ;\lsup{2}{\fR}) \cong O(D ,0;\fR)\oplus O(D ,0;\fR)$ is needed for the
classification.

\mysubsection{Defining equations}\label{s.la_def}
The defining equations for the classical LA's take the form,
\begin{subequations}
\begin{gather}
L^\dagger + L = 0
\label{eq.la} \\
\intertext{where,}
L \subset \widetilde{\fF}(D_+,D_-)\;,\quad\text{or}\quad
L \subset \overline{\fF}(D_+,D_-) 
\label{eq.la_ad}
\end{gather} 
\label{eq.la_def}

The adjoint ($\dagger$) is given by Eq.~\eqref{eq.tp_ad_def},
and involves the adjoint of $\fF$ and the metric $g$.
The adjoint of $\fF$ associated with each LA 
---  Reversion or Clifford Conjugation --- 
is indicated in Table~\ref{t.la}.

In the conventional classification of the LA's, 
skew-symmetric metrics are employed
for the symplectic groups.  In contrast, under the current classification
$g$ is always symmetric.
$g = I_{D_+,D_-}$, the pseudo-Euclidean metric, for $O(D_+,D_- ;\fR)$, $U(D_+,D_- ;\fC)$ and $U(D_+,D_- ;\fH)$;
the other LA's are independent of the signature of the metric, so we are free to choose $g = I$.

\begin{rem}
Eq.~\eqref{eq.la_def} defines the automorphism algebras for
$\widetilde{\fF}(D_+,D_-)$ and $\overline{\fF}(D_+,D_-)$,
\begin{equation}
L = 
\mathfrak{aut}\bigl(\fF(D_+,D_-),\dag\bigr)
\label{eq.la_aut}
\end{equation}
$L$ inherits the grading of $\fF$: this follows from Eq.~\eqref{eq.preserve}.
\begin{minorexmp}
Referring to table~\ref{t.la}, 
\begin{align*}
\mathfrak{aut}\bigl(\widetilde{\fH}(D_+,D_-)\bigr) & =  O(D_+ + D_-;\fH)\\
\mathfrak{aut}\bigl(\overline{\fH}(D_+,D_-)\bigr)  & =  U(D_+,D_- ;\fH)
\end{align*}
\end{minorexmp}
\end{rem}
\end{subequations}

\begin{exmp}[{\em Definition of ${U(D ;\fH'')}$}]
Consider the LA obtained from Eq.~\eqref{eq.la_def} with $\fF = \fH''\cong Cl(2,0;\fR)$, 
$g = I$ and
$\dagger = \text{Clifford Conjugate}$. $L$ may be expressed as a tensor product (Eq.~\eqref{eq.tp}),
\begin{equation}
L  =  L_1 \otimes 1 + L_{e_1} \otimes e_1  + L_{e_2} \otimes e_2  + L_{e_{12}} \otimes e_{12} 
\label{eq.ex1}
\end{equation}
where $L_i \subset \fR(D)$.  
Eq.'s~\eqref{eq.cliff},~\eqref{eq.r_ad}~and~\eqref{eq.la} then yield,
\begin{multline*}
0 = L + L^\dagger  = 
(L_1+L_1^t) \otimes 1 +(L_{e_1}- L_{e_1}^t) \otimes e_1  \\  + (L_{e_2}- L_{e_2}^t) \otimes e_2
+  (L_{e_{12}}- L_{e_{12}}^t) \otimes e_{12}
\end{multline*}
Consequently $L_1$ is skew-symmetric and  $L_{e_1}$, $L_{e_2}$ and $L_{e_{12}}$ are symmetric.
More generally, for any classical LA,
\[L_i^\dagger = \pm L_i\]
Let symmetric matrices be denoted by $S_i$ and skew-symmetric matrices by $A_i$.

A matrix-representation of $L$ may be constructed, using Eq.~\eqref{eq.ex1}
in conjunction with a matrix-representation of $\fF$.
An $\fR(2)$ representation of $\fF$ is given by,
\begin{equation}
1 = \bigl(\begin{smallmatrix} 1 & 0 \\ 0 & 1\end{smallmatrix}\bigr), 
\,e_1 = \bigl(\begin{smallmatrix} 1 & 0 \\ 0 & -1\end{smallmatrix}\bigr), 
\,e_2 = \bigl(\begin{smallmatrix} 0 & 1 \\ 1 & 0\end{smallmatrix}\bigr), 
\,e_{12} = \bigl(\begin{smallmatrix} 0 & 1 \\ -1 & 0\end{smallmatrix}\bigr)
\label{eq.cl2}
\end{equation}
leading to an $\fR(2D)$ representation of $L$,
\[
L = \left(\begin{array}{l|l}
A_1 + S_{e_1} & S_{e_2} + S_{e_{12}} \\
\hline
S_{e_2} - S_{e_{12}} & A_1 - S_{e_1}
\end{array}\right)
\]
This matches the representation of $Sp(2D;\fR)$~\cite{gilmore74}. 
\begin{numberedrem}
The  grading of $U(1;\fH'')$ and more generally of LA's with $D=1$ or $D_+ +D_-=1$ requires special consideration.  
The $1\times 1$ skew-symmetric matrices ($A_i$ above) are zero,  and this may reduce the grading.
For instance, in $O(1;\fH)$ only $L_{e_{12}}$ is non-zero.  The LA resides inside a subalgebra of $\fH$,
namely $(1,e_{12})$, and the subalgebra carries a $\ztwo^1$ grading, versus $\ztwo^2$ for $\fH$.
\label{r.reduced}
\end{numberedrem}
\end{exmp}

\subsubsection{Nomenclature}\label{ss.nom}
The traditional nomenclature for the symplectic algebras, $Sp(2D;\fR)$ and $Sp(2D;\fC)$,
is not appropriate for the description of these algebras in terms of
$\fF = \fH''$ and $\fC\otimes_\fR \fH''$: a new notation is desirable.

Since the classical LA's are automorphism algebras, they may be specified as
$\mathfrak{aut}\bigl(\widetilde{\fF}(D_+,D_-)\bigr)$ 
or
$\mathfrak{aut}\bigl(\overline{\fF}(D_+,D_-)\bigr)$
.

Alternatively, the designation of each LA as $O(D_+,D_- ;\fF)$ or $U(D_+,D_- ;\fF)$ is in
common use for the six $\fF$'s in Eq.~\eqref{eq.f2}~\cite{porteous95}, so to complete
the classification we need only consider the remaining two $\fF$'s: $\fH''$ and
$\fC\otimes_\fR \fH''$.

$U(D ;\fH'')$ is defined analogously to $U(D ,0;\fH)$; both LA's may also be
specified as 
$\mathfrak{aut}\bigl(\overline{\fF}(D,0)\bigr)$
, $\fF = \fH\;\text{or}\;\fH''$.

In order to obtain an alternative notation for 
$\mathfrak{aut}\bigl(\overline{\fC\otimes_\fR\fH''}(D,0)\bigr)$
,
let us first express the adjoint as a tensor product,
\[
\overline{\fC\otimes_\fR\fH''} = \widetilde{\fC}\otimes_\fR\overline{\fH''}
\]
Thus Clifford Conjugation of the tensor product is equivalent to Reversion of
$\fC$ tensored with Clifford Conjugation of $\fH''$.
Since 
$\mathfrak{aut}\bigl(\widetilde{\fC}(D,0)\bigr)\cong O(D ;\fC)$ 
and 
$\mathfrak{aut}\bigl(\overline{\fH''}(D,0)\bigr)\cong U(D ;\fH'')$
,
a logical but cumbersome designation for 
$\mathfrak{aut}\bigl(\overline{\fC\otimes_\fR\fH''}(D,0)\bigr)$
is
\mbox{$O_\fC U_{\fH''}(D;\fC\otimes_\fR \fH'')$}. \\
\begin{rem}
Reversion has no effect on $\fC\cong Cl(0,1;\fR)$, since $\tilde{1} = 1$ and 
$\tilde{e_1}=e_1$.  Consequently $O(D ;\fC)\cong \fC\otimes_\fR O(D ,0;\fR)$,
and $O_\fC U_{\fH''}(D;\fC\otimes_\fR \fH'') \cong \fC\otimes_\fR U(D ;\fH'')$.
\end{rem}

\mysection{Higher grading}\label{s.la_r}

\begin{table*} 
\begin{minipage}{0.95\textwidth}
\caption{
Equivalent adjoints for tensor products and higher Clifford algebras
\protect\footnote{The adjoint $\dag(a_+,a_-)$ is defined in Eq.~\protect\eqref{eq.anti}}
}\label{t.higher}
\begin{ruledtabular}
\begin{tabular}{ll} 
\multicolumn{1}{c}{Tensor Product} & 
\multicolumn{1}{c}{Higher Clifford Algebra}
\\
\hline
\\[-2.0ex]
$ \sotimes{n_+} Cl(2,0;\fR)^{\dag(2,0)}
  \sotimes{n_-} Cl(2,0;\fR)^{\dag(1,0)}
\otimes Cl(d_+,d_-;\fR)^{\dag(a_+,a_-)}
$
&
$Cl\bigl(d_+ + n_+ + n_- \bm{,}
d_- + n_+ + n_- + ; \fR
\bigr)^{\dag(a_+ + n_+ ,a_- + n_-)} 
$\\
$ \sotimes{n_+} Cl(2,0;\fR)^{\dag(2,0)}
  \sotimes{n_-} Cl(2,0;\fR)^{\dag(1,0)}
\otimes_\fR Cl(d;\fC)^{\dag(a)}
$
&
$ Cl\bigl(d + 2(n_+ + n_-) ; \fC
\bigr)^{\dag(a + n_+ + n_- )} 
$
\end{tabular}
\end{ruledtabular}
\end{minipage}
\end{table*}
\mysubsection{Tensor product of graded algebras}\label{ss.tp}
In Sec.~\ref{s.la_f} each classical LA was expressed as a subalgebra of $\fF(D)$ and consequently
inherited the grading of $\fF$.  $\fF(D)$ may be regarded as a tensor product of $\fR(D)$ and
$\fF$, (Eq.~\eqref{eq.tp}), and both components may carry a grading. 
The total grading is then the direct sum of the gradings for each term (cf Eq.~\eqref{eq.tp_grade}),
\[
\Gamma(\fF(D)) = \Gamma(\fR(D)) \oplus \Gamma(\fF)
\]

\begin{prop}
\[
\Gamma(\fF(D_+\,2^n,\,D_-\,2^n)) = 
\begin{cases}
\ztwo^{2n} \oplus \Gamma(\fF), & D_+ \ne D_- \\
\ztwo^{2(n+1)} \oplus \Gamma(\fF), & D_+ = D_- \\
\end{cases}
\]
and this grading is inherited by the classical LA's.
\end{prop}

\begin{proof}
$\fR(D_+\,2^n,\,D_-\,2^n)$ and its metric may be expressed as tensor products,
\begin{align}
\fR(D_+\,2^n,\,D_-\,2^n) &\cong  \fR(D_+,D_-) \sotimes{n} \fR(2) \label{eq.tp_rn}\\
I_{D_+\,2^n,\,D_-\,2^n} &\cong  I_{D_+,D_-} \sotimes{n} I_2
\label{eq.tp_metric}
\end{align}
where $\sotimes{n} \fR(2)$ denotes the tensor product of $n$ copies of $\fR(2)$.

$\fR(2)$ equipped with matrix transpose 
is isomorphic to $Cl(2,0;\fR)$ equipped with Reversion (cf Eq.~\eqref{eq.cl2}),
\begin{equation}
\bigl(\fR(2),\text{Transpose}\bigr) \cong \bigl(Cl(2,0;\fR),\text{Reversion}\bigr)
\label{eq.cl2_cong}
\end{equation}
Substituting Eq.'s~\eqref{eq.tp_rn}---\eqref{eq.cl2_cong} into Eq.~\eqref{eq.tp_ad_def} yields,
\begin{subequations}
\begin{equation}
\fF(D_+\,2^n,\,D_-\,2^n)  \cong  \fR(D_+,D_-) \sotimes{n} Cl(2,0;\fR) \otimes_\fR \fF \label{eq.tp_f} 
\end{equation}
\vspace*{-2ex}
\begin{multline}
 \bigl(\fR(D_+,D_-)  \sotimes{n} Cl(2,0;\fR) \otimes_\fR \fF\bigr)^\dag  = \\ 
g^{-1} \fR(D_+,D_-)^t g \sotimes{n} \widetilde{Cl(2,0;\fR)} \otimes_\fR \fF^\dag
\label{eq.tp_f_dag}  
\end{multline}
where,
\begin{equation}
g  =  I_{D_+,D_-} \label{eq.small_metric}
\end{equation}
\label{eq.tp_f_def}
\end{subequations}
$Cl(2,0;\fR)$ and $\fF$ carry $\ztwo^2$ and $\Gamma(\fF)$ gradings which are invariant under the adjoint,
and it follows from Eq.~\eqref{eq.tp_f_def} that $\fF(D_+\,2^n,\,D_-\,2^n)$ carries a $\ztwo^{2n} + \Gamma(\fF)$
grading which is also invariant under the adjoint.  Consequently the LA (Eq.~\eqref{eq.la_def})
inherits the grading.

For the special case $D_+ = D_-$, $\fR(D_+,D_+)$ may be further factorized as
\begin{align*}
\fR(D_+,D_+) & \cong  \fR(D_+,0)\otimes \fR(1,1) \\
I_{D_+,D_+}  & \cong  I_{D_+,0}\otimes I_{1,1} 
\end{align*}
$\fR(1,1)$ equipped with the adjoint $I_{1,1}^{-1} \fR(1,1)^t I_{1,1}$ is isomorphic to
$Cl(2,0;\fR)$ equipped with $\dag(1,0)$,
\begin{subequations}
\begin{align}
\bigl(\fR(1,1),\dag\bigr) & \cong \bigl(Cl(2,0;\fR),\dag(1,0)\bigr) \nonumber\\
\fR(D_+,D_+) & \cong  \fR(D_+,0)\otimes Cl(2,0;\fR) \\
\bigl(\fR(D_+,0)\otimes Cl(2,0;\fR)\bigr)^\dag & =   \fR(D_+,0)^t \otimes Cl(2,0;\fR)^{\dag(1,0)}
\end{align}
\label{eq.sc}
\end{subequations}
and the attendant grading is $\ztwo^{2(n+1)} \oplus \Gamma(\fF)$.

\end{proof}

\begin{minorexmp}
$U(D;\,\lsup{2}{\fH})$ has a $\ztwo^3$ grading, and $U(D\,2^n;\,\lsup{2}{\fH})$ has a $\ztwo^{3+2n}$ grading.
\end{minorexmp}

\begin{rem}
The grading of LA's associated with $\fF(D_+\,2^n,\,D_-\,2^n)$
may be reduced for the special case $D_+ +D_- =1$, cf Remark~\ref{r.reduced}.
\end{rem}

\mysubsection{Higher Clifford algebras}\label{ss.hca}
  
In Sec.~\ref{ss.tp} the grading of $\fF(D_+\,2^n,\,D_-\,2^n)$ was exposed by expressing the algebra as a
tensor product of graded algebras (Eq.~\eqref{eq.tp_f_def}).  In this section the 
tensor product is expressed in terms of a higher Clifford algebra which plays a role analogous 
to $\fF$ in Sec.~\ref{s.la_f}.

$\fF(D_+\,2^n,\,D_-\,2^n)$ is first expressed as,
\begin{subequations}
\begin{gather}
\fF(D_+\,2^n,\,D_-\,2^n) \cong 
\begin{cases}
\fA(D_+ ,D_- ), & D_+ \ne D_- \\
\fA(D_+ ,0), & D_+ = D_- 
\end{cases}
\label{eq.fa} \\
\intertext{where $\fA \supset \fF$ is a Clifford algebra with, }
\Gamma(\fA) = 
\begin{cases}
\ztwo^{2n} \oplus \Gamma(\fF), & D_+ \ne D_- \\
\ztwo^{2(n+1)} \oplus \Gamma(\fF), & D_+ = D_- \\
\end{cases}
\end{gather} 
We subsequently determine $(m_+,m_-)$ such that,
\begin{multline}
\bigl(\fF(D_+\,2^n,\,D_-\,2^n),\dag\bigr) \cong \\
\begin{cases}
\bigl(\fA(D_+,D_-),\dag(m_+,m_-)\bigr), &  D_+ \ne D_- \\
\bigl(\fA(D_+,0),\dag(m_+,m_-)\bigr), &  D_+ = D_- 
\end{cases}
\label{eq.fa_ad}
\end{multline}
\label{eq.fa_def}
\end{subequations}
It follows that the LA inherits the grading of $\fA$ and that the LA may be defined directly in terms of
a Clifford algebra with a manifest grading,
\begin{equation}
L = 
\begin{cases}
\mathfrak{aut}\bigl(\fA(D_+,D_-),\dag(m_+,m_-)\bigr), & D_+ \ne D_- \\
\mathfrak{aut}\bigl(\fA(D_+,0),\dag(m_+,m_-)\bigr), & D_+ = D_- 
\end{cases}
\end{equation}

\begin{lem} \label{lem.t_alg}
\begin{subequations}
\begin{align}
\sotimes{n} Cl(2,0;\fR) \otimes Cl(d_+,d_-;\fR)
& \cong   Cl(d_+ +n,d_- +n;\fR)
\label{eq.t_alg_r} \\
\sotimes{n} Cl(2,0;\fR) \otimes_\fR Cl(d;\fC)
& \cong   Cl(d+2n;\fC)
\label{eq.t_alg_c} 
\end{align}
\end{subequations}
\end{lem}
\begin{proof}
A standard expression for tensor products involving $Cl(2,0;\fR)$ is~\cite{porteous95},
\begin{equation}
Cl(2,0;\fR) \otimes Cl(d_+,d_-;\fR) \cong Cl(d_+ +1,d_- +1;\fR)
\label{eq.plus1}
\end{equation}
Eq.~\eqref{eq.t_alg_r} follows by induction.
Eq.~\eqref{eq.t_alg_c} follows upon tensoring Eq~\eqref{eq.t_alg_r} with $\fC$ and
applying Eq.~\eqref{eq.complex}.
\end{proof}

\begin{lem} \label{lem.t_ad}
The adjoint for each tensor product in Table~\ref{t.higher} is equivalent to the adjoint of the
higher Clifford algebra in the same row.
\end{lem}

\begin{proof}
Let us first consider the real Clifford algebras.  Let 
$e_i,\; i \le d_+ +d_-,$ denote basis elements for $Cl(d_+,d_-;\fR)$
, and
$E_i,\; i \le 2,$ denote basis elements for
$Cl(2,0;\fR)$ 
.

The $(d_+ +d_- +2)$ basis elements for $Cl(d_+ +1,d_- +1;\fR)$ are~\cite{brauer35},
\begin{equation*}
\left\{
E_1\otimes e_i,\;E_2\otimes 1,\;E_{12}\otimes 1
\right\}
\end{equation*}

Applying the tensor-product adjoint yields,
\begin{multline*}
\left\{
E_1^{\dag(2,0)}
\otimes 
e_i^{\dag(a_+,a_-)}
,\;
E_2^{\dag(2,0)}
\otimes 
1
,\;
E_{12}^{\dag(2,0)}
\otimes 
1
\right\}  = \\  
\left\{
{E_1}
\otimes 
e_i^{\dag(a_+,a_-)}
,\;
{E_2}
\otimes 
1
,\;
-{E_{12}}
\otimes 
1
\right\}  
\end{multline*}
$E_2\otimes 1$ squares to $+1$ and does not change sign under the adjoint,
hence the corresponding adjoint of $Cl(d_+ +1,d_- +1;\fR)$ is $\dag(a_+ + 1,a_-)$.
Replacing $E_i^{\dag(2,0)}$ with $E_i^{\dag(1,0)}$ yields $\dag(a_+,a_- + 1)$,
and the first row of Table~\ref{t.higher} follows by induction on $n_+$ and $n_-$.
The second row follows upon tensoring the first row with $\fC$ and
applying Eq.~\eqref{eq.complex}.
\end{proof}

\begin{prop}\mbox{ }\newline
For $ \fF \cong Cl(\left.d_\fF\right._+,\left.d_\fF\right._-;\fR) $,
\begin{multline*}
\bigl(\fF(D_+\,2^n,\,D_-\,2^n),\dag(\left.a_\fF\right._+,\left.a_\fF\right._-)\bigr) \cong \\
\begin{cases}
\bigl(\fA(D_+,\,D_-),{\dag(\left.a_\fF\right._+ + n,\left.a_\fF\right._-)}\bigr),
& D_+ \ne D_- \\
\bigl(\fA(D_+,\,0),{\dag(\left.a_\fF\right._+ + n,\left.a_\fF\right._- + 1)}\bigr),
& D_+ = D_- \\
\end{cases}
\end{multline*}
where,
\[
\fA = 
\begin{cases}
Cl(\left.d_\fF\right._+ + n,\left.d_\fF\right._- + n;\fR),  & D_+ \ne D_- \\
Cl(\left.d_\fF\right._+ + n + 1,\left.d_\fF\right._- + n + 1;\fR),  & D_+ = D_-
\end{cases}
\]
For $\fF = \,\lsup{2}{\fC} \cong Cl(1;\fC)$,
\begin{gather*}
\bigl(\,\lsup{2}{\fC}(D\,2^n),\dag(a_\fF)\bigr) \cong \bigl(\fA(D),{\dag(a_\fF + n)}\bigr) \\
\intertext{where}
\fA = Cl(1 + 2n,\fC) 
\end{gather*} 
\label{p.pad}
\end{prop}

\begin{proof}
Eq.~\eqref{eq.fa} may be proved by applying Lemma~\ref{lem.t_alg} to Eq.~\eqref{eq.tp_f}, where
$\fF$ and $\fA$ are as defined in Prop.~\ref{p.pad}.

The remaining task is to
establish the connection between the adjoints of $\fF(D_+\,2^n,\,D_-\,2^n)$ and
$\fA(D_+,D_-)$ which arise in the proposition.

Setting, 
\begin{align*}
(d_+,d_-) & = (\left.d_\fF\right._+,  \left.d_\fF\right._-),\quad
(a_+,a_-) = (\left.a_\fF\right._+,  \left.a_\fF\right._-),\\
(n_+,n_-) & = (n,0)
\end{align*}
in the first row of Table~\ref{t.higher}
and then comparing with Eq.~\eqref{eq.tp_f_def} 
with $(D_+,D_-)$ set to $(1,0)$ yields,
\[
\bigl(
\fF(2^n),
\dag(\left.a_\fF\right._+,  \left.a_\fF\right._-)
\bigr) 
\cong 
\bigl(
\fA,\\
\dag(\left.a_\fF\right._+ + n,  \left.a_\fF\right._-)
\bigr)
\]
This establishes the connection between the adjoints for the real case, and
Prop.~\ref{p.pad} follows
upon taking the tensor product of this equation with $\fR(D_+,D_-)$. 

The complex case follows upon taking the tensor product with $\fC$ and applying Eq.~\eqref{eq.complex}.
The special case, $D_+ = D_-$, also involves Eq.~\eqref{eq.sc} and requires setting $(n_+,n_-) = (n,1)$
in Table~\ref{t.higher}.
\end{proof}

\begin{minorexmp}
\begin{gather*}
\Bigl(\, \lsup{2}{\fH}(D\,2^n),\text{Reversion}\Bigr)  \cong \bigl(\fA(D),\dag(n,3)\bigr) \\
\intertext{where}
\fA  = Cl(n,n+3;\fR)\\
\intertext{Hence,}
U(D\,2^n;\,\lsup{2}{\fH})  = \mathfrak{aut}\bigl(\fA(D),\dag(n,3)\bigr) 
\end{gather*}
and both $\fA$ and $U(D\,2^n;\,\lsup{2}{\fH})$ have a $\ztwo^{3+2n}$ grading.
\end{minorexmp}

\begin{numberedrem}
Prop.~\ref{p.pad} employs non-standard adjoints of $\fA$, ie adjoints other than Clifford Conjugation and
Reversion.  In some cases it is possible to replace a non-standard adjoint with a standard adjoint
by exploiting the periodicity properties of the Clifford algebras:
\begin{exmp}
Setting $\fF=\fR$, $n=D_+ =1$ and $D_- =0$ in Prop.~\ref{p.pad} yields,
\[
\bigl(\fR(2),\text{Transpose}\bigr) \cong \bigl(Cl(1,1;\fR),\dag(1,0)\bigr)
\]
An alternative involving a standard adjoint is provided by $Cl(2,0;\fR)$ equipped with Reversion,
\begin{align*}
\bigl(\fR(2),\text{Transpose}\bigr) & \cong \bigl(Cl(1,1;\fR),\dag(1,0)\bigr) \\
& \cong \bigl(Cl(2,0;\fR),\text{Reversion}\bigr)
\end{align*}
\end{exmp}
\begin{exmp}
Setting $\fF=\fR$, $n=2$, $D_+ =1$ and $D_- =0$ in Prop.~\ref{p.pad} yields,
\[
\bigl(\fR(4),\text{Transpose}\bigr) \cong \bigl(Cl(2,2;\fR),\dag(2,0)\bigr)
\]
In this case no standard adjoint can be employed, because
$Cl(2,2;\fR)$ is not isomorphic to $Cl(4,0;\fR)$ or $Cl(0,4;\fR)$,
\[
Cl(2,2;\fR)\cong \fR(4) \ncong \fH(2) \cong Cl(4,0;\fR) \cong(0,4;\fR)
\]
\end{exmp}
\begin{exmp}
For $\fF = \lsup{2}{\fC} \cong  Cl(1;\fC)$ equipped with Clifford Conjugation, it is always possible to use a 
standard adjoint,
\begin{gather}
\overline{\lsup{2}{\fC}(D\,2^n)}\cong
\begin{cases}
\overline{\fA(D)}, & n\;\text{even} \\[1.5ex]
\widetilde{\fA(D)}, & n\;\text{odd} 
\end{cases}
\\ \intertext{where} \fA = Cl(1 + 2n,\fC)
\end{gather}
This result is implicit in Porteous's survey of the periodicity properties of the
real and complex Clifford algebras~\cite{porteous95}.
\end{exmp}
\end{numberedrem}

\mysubsection{Non-tensorial grading}
The grading described in Sec's.~\ref{ss.tp} and~\ref{ss.hca} reflects the tensorial structure of the
Lie algebras.  It is also possible to impose a $\ztwo^n$ grading that is not associated with
a tensorial structure.  For example, a $\ztwo^n$ grading can be imposed on a vector space or a matrix.  
In this section we briefly describe such a grading.

In Sec.~\ref{ss.hca} each classical LA was expressed as a subalgebra of $\fA(D)$ or $\fA(D_+,D_-)$ and consequently
inherited the grading of $\fA$.  $\fA(D)$ may be regarded as a tensor product of $\fR(D)$ with
$\fA$, and $\fR(D)$ may carry a non-tensorial grading.
The total grading is then the direct sum of the gradings for each term (cf Eq.~\eqref{eq.tp_grade}),
\[
\Gamma(\fA(D)) = \Gamma(\fR(D)) \oplus \Gamma(\fA)
\]
The $(D_+,D_-)$ metric does not affect the non-tensorial grading, and hence $\fR(D_+,D_-)$ is equivalent
to $\fR(D_+ + D_-)$ for grading purposes.

A $\ztwo^1$  grading may be imposed by partitioning $\fR(D)$ into diagonal and off-diagonal blocks.
The diagonal blocks must be square but the dimensions are otherwise arbitrary,
\begin{gather*}
\fR(D) = \fR(D)_0 + \fR(D)_1\;, \\
\end{gather*}
where,
\begin{scriptsize}
\begin{gather*}
\fR(D) = 
\left[
\begin{array}{c|cc}
\fR(D)_0 & \fR(D)_1 &\hphantom{xx} \\[8.0ex]
\hline
\fR(D)_1 & \fR(D)_0 & \\[12.8ex]
\end{array}
\right]
\end{gather*}
\end{scriptsize}
The grading is well-behaved under multiplication and invariant under transpose. It is also invariant
under the  adjoint (Eq.~\eqref{eq.r_ad}), because the metric is diagonal (pseudo-Euclidean) and hence
does not mix matrix elements with different grades (Sec.~\ref{s.la_def}).  The grading is not compatible with
a skew-symmetric metric, but in the present work only pseudo-Euclidean metrics are employed.
The arbitrary subdivision may be repeated to yield a $\ztwo^2$ grading,
\[
\fR(D) = \fR(D)_{00} + \fR(D)_{01} + \fR(D)_{10} +  \fR(D)_{11} 
\]
where,
\begin{scriptsize}
\begin{multline*}
\fR(D) = \\
\left[
\begin{array}{c|cccccccccccccccc|ccc}
\fR(D)_{00} & \fR(D)_{01} &&&& \vline\vline & \fR(D)_{10} &&&&&&&&&&   & \fR(D)_{11}   & & \\[9.5ex]
\hline
\fR(D)_{01} & \fR(D)_{00}   &&&&  \vline\vline & \fR(D)_{11} &&&&&&&&&&  & \fR(D)_{10} & & \\[15.3ex]
\hline  \\[-2.70ex]\hline
\fR(D)_{10} & \fR(D)_{11}  &&&&  \vline\vline & \fR(D)_{00} &&&&&&&&&&  & \fR(D)_{01}  & &  \\[31.0ex]
\hline
\fR(D)_{11} & \fR(D)_{10}  &&&&  \vline\vline & \fR(D)_{01} &&&&&&&&&&  & \fR(D)_{00}  & &  \\[13.4ex]
\end{array}
\right]
\end{multline*}
\end{scriptsize}
Subdividing $s$ times yields a $\ztwo^s$ grading.  Each subdivision of the
$D\times D$ matrix doubles the number of blocks in each row,  
and hence the non-tensorial grading is constrained by,
\[
2^s \le D
\]
For comparison, the tensorial subdivision with the same number of blocks, $\sotimes{s}\fR(2)\cong
Cl(s,s;\fR)$, has double the grading, $\ztwo^{2s}$.

\mysection{Conclusion}\label{s.conclusion}
To illuminate the $\ztwo^n$ grading of the classical Lie algebras, the
conventional classification in terms of matrix algebras over  two, three or six Clifford algebras 
has been expanded to include 
a total of eight Clifford algebras.  In so doing, the symplectic algebras have been
placed on an equal footing with the orthogonal and unitary algebras.  In particular, all the
Lie algebras have been expressed in terms of pseudo-Euclidean metrics.
The eight Clifford algebras and associated classical Lie algebras have gradings ranging from zero for 
$O(N;\fR)$ to $\ztwo^3$ for $Sp(2N;\fC) \cong \mathfrak{aut}(\overline{Cl(3,0;\fR)}(N))$ and 
$U\lstar(2N;\fC)\cong \mathfrak{aut}(\widetilde{Cl(0,3;\fR)}(N))$, as per Table~\ref{t.la}.  

Lie algebras with matrix dimensions divisible by a power of two have been recast in terms of matrix 
algebras over higher Clifford algebras with a higher grading.  
For example, the $\ztwo^{2n+3}$ grading of $U(m\, 2^n, m\,2^n;\fC)$ becomes apparent when the Lie
algebra is expressed in terms of a matrix algebra over
$Cl(n+1,n+2;\fR)$.
The defining equations for the Lie algebras involve a matrix adjoint operator, and this generally corresponds to a
non-standard adjoint of the higher Clifford algebras, ie an adjoint other than Clifford Reversion or
Clifford Conjugation.
It is hoped that the description of Lie algebras in terms of matrix algebras over higher Clifford algebras will
facilitate the description of the associated Lie color algebras.

\acknowledgments
The author is grateful for enlightening discussions with Haisheng Li and Gabor Toth.

This work was supported by Research Corporation through a Cottrel
College Science grant, No. CC5459.

\bibliography{}
\end{document}